\documentclass[superscriptaddress]{revtex4}

\usepackage{color,psfig,graphics,psfrag,amsmath,amssymb,amsfonts,rotating}

\newcommand{\eps}{\varepsilon}

\begin{document}

\title{On the Feasibility of Portfolio Optimization under Expected Shortfall}

\author{Stefano Ciliberti}
\affiliation{CNRS; Univ. Paris Sud, UMR8626, LPTMS, ORSAY CEDEX, F-91405}

\author{Imre Kondor}
\affiliation{Collegium Budapest, 1014 Budapest, Szentháromság u.~2}

\author{Marc M\'ezard}
\affiliation{CNRS; Univ. Paris Sud, UMR8626, LPTMS, ORSAY CEDEX, F-91405}


\begin{abstract}
We address the problem of portfolio optimization under the simplest coherent
risk measure, i.e. the expected shortfall. As it is well known, one can map
this problem into a linear programming setting. For some values of the
external parameters, when the available time series is too short, the
portfolio optimization is ill posed because it leads to unbounded positions,
infinitely short on some assets and infinitely long on some others. As first 
observed by Kondor and coworkers, this phenomenon is actually a phase transition. 
We investigate the nature of this transition by means of a replica approach.
\end{abstract}

\pacs{ }
 
\maketitle

\section{Introduction \label{sec:intro}}

Among the several risk measures existing in the context of portfolio
optimization, expected shortfall (ES) has certainly gained increasing 
popularity in recent years. In several practical applications, ES is 
starting to replace the classical Value-at-Risk (VaR). There are a number of 
reasons for this. For a given threshold probability $\beta$, the VaR is defined so that
with probability $\beta$ the loss will be smaller than VaR. This definition
only gives the minimum loss one can reasonably expect but does not tell
anything about the typical value of that loss that can be measured by the
 \emph{conditional} value-at-risk (CVaR, which is the same
as ES for continuous distributions that we consider here
\footnote{see~\cite{acerbi02} for the subtleties related to a discrete
distribution.} ).We will be more precise on these definitions below. The
point we want to stress here is that the VaR measure, lacking the mandatory
properties of subadditivity and convexity, is not
\emph{coherent}~\cite{artzner99}.  This means that summing VaR's of
individual portfolios will not necessarily produce an upper bound for the
VaR of the combined portfolio, thus contradicting the holy principle of
diversification in finance. A nice practical example of the inconsistency of
VaR in credit portfolio management is reported in Ref.~\onlinecite{frey02}.
On the other hand, it has been shown~\cite{acerbi02} that ES is a coherent
measure with interesting properties~\cite{pflug00}.  Moreover, the
optimization of ES can be reduced to linear programming~\cite{rockafellar00}
(which allows for a fast implementation) and leads to a good estimate for
the VaR as a byproduct of the minimization process. To summarize, the
intuitive and simple character, together with the mathematical properties
(coherence) and the fast algorithmic implementation (linear programming),
are the main reasons behind the growing importance of ES as a risk measure.

In this paper, we will focus on the feasibility of the portfolio
optimization problem under the ES measure of risk. The control parameters of
this problem are ($i$) the imposed threshold in probability, $\beta$, and
($ii$) the ratio $N/T$ between the number $N$ of financial assets making up
the portfolio and the time series length $T$ used to sample the probability
distribution of returns. (It is curious that, albeit trivial, the scaling in
$N/T$ had not been explicitly pointed out before~\cite{pafka02}.) It has
been discovered in \cite{kondor05} that, for certain values of these
parameters, the optimization problem does not have a finite solution
because, even if convex, it is not bounded from below.  Extended numerical
simulations allowed these authors to determine the feasibility map of the
problem. Here, in order to better understand the root of the problem and to
study the transition from a feasible regime to an unfeasible one
(corresponding to an ill-posed minimization problem) we address the same
problem from an analytical point of view.

The paper is organized as follows. In Section \ref{sec:opt} we briefly
recall the basic definitions of $\beta$-VaR and $\beta$-CVaR and we show how
the portfolio optimization problem can be reduced to linear programming. We
introduce a ``cost function'' to be minimized under linear constraints and
we discuss the rationale for a statistical mechanics approach. In Section
\ref{sec:replica} we solve the problem of optimizing large portfolios under
ES using the replica approach. Our results and the comparison with
numerics are reported in Section \ref{sec:phasediag}, and our conclusions
are summarized in Section \ref{sec:conclu}.

\section{The optimization problem \label{sec:opt}}

\begin{figure}
  \psfrag{b}[][][3]{$\quad \beta$}
  \psfrag{a}[][][3]{$\alpha$}
  \psfrag{bVaR}[][][3]{$\beta$-VaR$({\bf w})$}
  \psfrag{P(w;a)}[][][3]{$\mathcal{P}_<({\bf w};\alpha)$}
  \begin{center}
    \includegraphics[angle=-90,width=.6\textwidth]{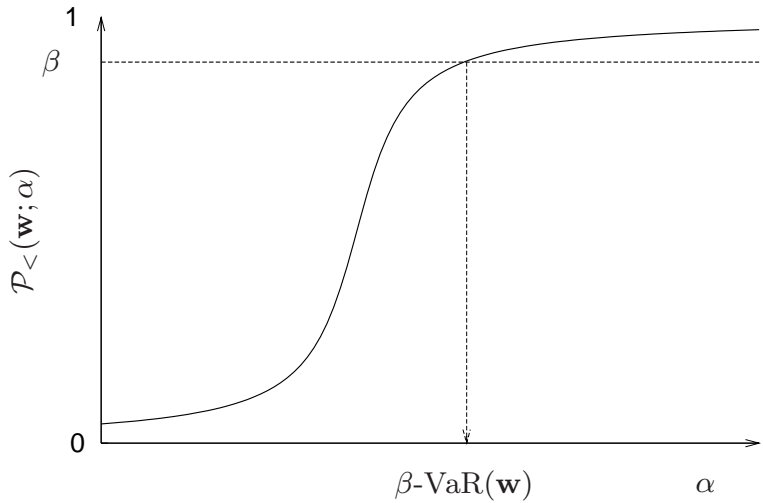}
  \end{center}
  \caption{Schematic representation of the VaR measure of risk.
    $\mathcal{P}_< ({\bf w})$ is the probability of a loss associated to the
    portfolio ${\bf w}$ being smaller than $\alpha$. The conditional VaR
    $\beta$-CVaR (or ES) is the average loss when this is constrained to be
    greater than the $\beta$-VaR.}
\label{fig:var}
\end{figure}

We consider a portfolio of $N$ financial instruments ${\bf w}=\{w_1, \ldots
w_N\}$, where $w_i$ is the position of asset $i$. The global
budget constraint fixes the sum of these numbers: we impose for example
\begin{eqnarray}
  \sum_{i=1}^N w_i = N \ .
  \label{eq:budget}
\end{eqnarray}
We do not stipulate any constraint on short selling, so that $w_i$ can be
any negative or positive number. This is, of course, irrealistic for
liquidity reasons, but considering this case allows us to show up the
essence of the phenomenon. If we imposed a constraint that would render the
domain of the $w_i$ bounded (such as a ban on short selling), this would
evidently prevent the weights from diverging, but a vestige of the
transition would still remain in the form of large, though finite,
fluctuations of the weights, and in a large number of them sticking to the
``walls'' of the domain.

We denote the returns on the assets by ${\bf x}=\{ x_{1}, x_{2}, \ldots
x_{N}\}$ , and we will assume an underlying normal distribution $p({\bf x})
\sim \prod_{i}\exp(-Nx^2_{i}/2)$.  The corresponding \emph{loss} is
$\ell({\bf w}|{\bf x})= -\sum_{i=1}^N w_i x_{i} $, and the probability of
that loss being smaller than a given threshold $\alpha$ is
\begin{equation}
  \mathcal{P}_<({\bf w}, \alpha ) = \int d{\bf x} \ p({\bf x})
  \theta\big(\alpha-\ell({\bf w}|{\bf x})\big) \ ,
\end{equation}
where $\theta(\cdot)$ is the Heaviside step function, equal to 1 if its
argument is positive and 0 otherwise. The $\beta$-VaR of this
portfolio is formally defined by
\begin{equation}
  \beta\textrm{-VaR} ({\bf w}) = \min \{ \alpha : \
  \mathcal{P}_<({\bf w},\alpha)\ge \beta\} \ ,
\end{equation}
(see Fig.~\ref{fig:var}), while the CVaR (or ES, in this case) associated
with the same portfolio is the average loss on the tail of the
distribution,
\begin{equation}
  \beta\textrm{-CVaR}({\bf w}) = 
  \frac
      {\displaystyle 
        \int d{\bf x} \ p({\bf x}) 
        \ell({\bf w}|{\bf x})
        \theta\big(\ell({\bf w}|{\bf x})-\beta\textrm{-VaR} ({\bf w})\big)}
      {\displaystyle 
        \int d{\bf x}\  p({\bf x}) 
        \theta\big(\ell({\bf w}|{\bf x})-\beta\textrm{-VaR}({\bf w})\big)}
      =
      \frac{1}{1-\beta}
        \int d{\bf x} \ p({\bf x}) 
        \ell({\bf w}|{\bf x})
        \theta\big(\ell({\bf w}|{\bf x})-\beta\textrm{-VaR} ({\bf w})\big)      
        \ .
        \label{eq:betacvar}
\end{equation}
The threshold $\beta$ then represents a confidence level.  In practice, the
typical values of $\beta$ which one considers are $\beta=0.90, 0.95,$ and $
0.99$, but we will address the problem for any $\beta \in [0,1]$.  What is
usually called ``exceedance probability'' in some previous literature would
correspond here to $(1-\beta)$.

As mentioned in the introduction, the ES measure can be obtained from a
variational principle~\cite{rockafellar00}. The minimization of a properly
chosen objective function leads directly to (\ref{eq:betacvar}):
\begin{eqnarray}
  \beta\textrm{-CVaR}({\bf w}) & = & \min_{v} F_\beta({\bf w},v) \ ,\\
  F_\beta({\bf w},v) & \equiv & 
  v + (1-\beta)^{-1}  \int d{\bf x}\  p({\bf x}) 
  \big[ \ell({\bf w}|{\bf x}) - v \big]^+\ .
  \label{eq:objective}
\end{eqnarray}
Here, $[a]^+ \equiv (a+|a|)/2$. The external parameter $v$ over which one
has to minimize is claimed to be relevant in itself~\cite{rockafellar00},
since its optimal value may represent a good estimate for the actual
value-at-risk of the portfolio. We will come back to this point as we
discuss our results.  We stress here that minimizing (\ref{eq:objective})
over ${\bf w}$ \emph{and} $v$ is equivalent to optimizing
(\ref{eq:betacvar}) over the portfolio vectors ${\bf w}$.

Of course, in practical cases the probability distribution of the loss is
not known and must be inferred from the past data. In other words, we need
an ``in-sample'' estimate of the integral in (\ref{eq:objective}), which
would turn a well posed (but useless) optimization problem into a practical
approach.  We thus approximate the integral by sampling the probability
distributions of returns. For a given time series ${\bf x}^{(1)}, \ldots
{\bf x}^{(T)}$, our objective function becomes simply
\begin{equation}
  \hat F_\beta({\bf w},v) = v + \frac{1}{(1-\beta)T}
  \sum_{\tau=1}^T
  \big[ \ell({\bf w}|{\bf x}^{(\tau)}) - v \big]^+ 
  =
  v + \frac 1{(1-\beta) T} \sum_{\tau=1}^T 
  \left[ -v-\sum_{i=1}^N w_i x_{i\tau} \right]^+ 
  \ ,
  \label{eq:objective2}
\end{equation}
where we denote by $x_{i\tau}$ the return of asset $i$ at time $\tau$.
Optimizing this risk measure is the same as the following linear programming
problem:
\begin{itemize}
  \item
    given one data sample, i.e. a matrix $x_{i\tau}$, $i=1,\ldots N$,
    $\tau=1,\ldots T$,
  \item
    minimize the \emph{cost function}
    \begin{equation}
      E_\beta\big[v,\{w_i\},\{ u_\tau\};\{x_{i\tau}\}\big]
      = (1-\beta)Tv+\sum_{\tau=1}^T u_\tau \ ,
      \label{eq:ebeta}
    \end{equation}
  \item
    over the $(N+T+1)$ variables ${\bf Y}\equiv \{w_1,\ldots w_N,u_1,\ldots
    u_T\,v\}$,
  \item
    under the $(2T+1)$ constraints
    \begin{equation}
      u_\tau \ge  0 \ ,\quad   
      u_\tau + v + \sum_{i=1}^N x_{i\tau} w_i \ge 0 \quad
      \forall \tau\ , \ \ \textrm{and}\quad \sum_{i=1}^N w_i = N \ .
      \label{eq:constraints}
    \end{equation}
\end{itemize}
Since we allow short positions, not all the $w_i$ are positive, which makes
this problem different from standard linear programming. To keep the problem
tractable, we impose the condition that $w_i \ge -W$, where $W$ is a very
large cutoff, and the optimization problem will be said to be ill-defined if
its solution does not converge to a finite limit when $W\to\infty$. It is
now clear why constraining all the $w_i$ to be non-negative would eliminate
the feasibility problem: a finite solution will always exists because the
weights are by definition bounded, the worst case being an optimal portfolio
with only one non-zero weight taking care of the total budget. The control
parameters that govern the problem are the threshold $\beta$ and the ratio
$N/T$ of assets to data points. The resulting ``phase diagram'' is then a
line in the $\beta$-$N/T$ plane separating a region in which, with high
probability, the minimization problem is not bounded and thus does not admit
a finite solution, and another region in which a finite solution
exists. These statements are non-deterministic because of the intrinsic
probabilistic nature of the returns. We will address this minimization
problem in the non-trivial limit where $T\to\infty$, $N\to\infty$, while
$N/T$ stays finite. In this ``thermodynamic'' limit, we shall assume that
extensive quantities (like the average loss of the optimal portfolio,
i.e. the minimum cost function) do not fluctuate, namely that their
probability distribution is concentrated around the mean value.  This
``self-averaging'' property has been proven for a wide range of similar
statistical mechanics models~\cite{selfav}.  Then, we will be interested in
the average value of the min of (\ref{eq:ebeta}) over the distribution of
returns. Given the similarity of portfolio optimization with the statistical
physics of disordered systems, this problem can be addressed analytically by
means of a replica approach~\cite{mepavi}.

\section{The replica approach \label{sec:replica}}

We consider one given sample, i.e. a given history of returns $x_{i\tau}$
drawn from the distribution
\begin{equation}
  p(\{x_{i\tau}\}) \sim \prod_{i\tau} e^{-N x^2_{i\tau}/2} \ .
  \label{eq:probpx}
\end{equation}
In order to compute the minimal cost, we introduce the partition function at
inverse temperature $\gamma$. Recalling that ${\bf Y}$ is the set of all
variables, the partition function at inverse temperature $\gamma$ is defined as
\begin{eqnarray}
  Z_\gamma[\{x_{i\tau}\}] & = & 
  \int_V d{\bf Y} 
  \exp\Big[-\gamma E_\beta[{\bf Y};\{ x_{i\tau}\}]\Big]
  \label{eq:part0}
\end{eqnarray}
where $V$ is the convex polytope defined by (\ref{eq:constraints}). The
intensive minimal cost corresponding to this sample is then computed as
\begin{equation}
  \eps[\{x_{i\tau}\}] = \lim_{N\to\infty} \frac {\min E[\{ x_{i\tau} \}] }{N}
  = \lim_{N\to\infty}\lim_{\gamma\to\infty} \frac{-1}{N\gamma}
  \log Z_\gamma[\{x_{i\tau}\}] \ .
  \label{eq:epsxit}
\end{equation}
Actually, we are interested in the average value of this quantity over the
choice of the sample. Equation (\ref{eq:epsxit}) tells us that the average
minimum cost depends on the average of the \textsl{logarithm} of $Z$. This
difficulty is usually circumvented by means of the so called ``replica
trick'': one computes the average of $Z^n$, where $n$ is an integer, and
then the average of the logarithm is obtained by
\begin{equation}
  \overline{\log Z} 
  = \lim_{n\to 0} \frac{\partial\overline{ Z^n}}{\partial n} \ ,
  \label{eq:replicatrick}
\end{equation}
thus assuming that $Z^n$ can be analytically continued to real values of
$n$. The overline stands for an average over different samples, i.e. over
the probability distribution (\ref{eq:probpx}). This technique has a long
history in the physics of spin glasses~\cite{mepavi}: the proof that it leads to
the correct solution has been obtained~\cite{talagrand} recently.

The partition function (\ref{eq:part0}) can be written more explicitly as
\begin{eqnarray}
  Z_\gamma[\{x_{i\tau}\}] & = & 
  \int_{-\infty}^{+\infty}dv 
  \int_{0}^{+\infty} \prod_{\tau=1}^T du_\tau
  \int_{-\infty}^{+\infty} \prod_{i=1}^N dw_i
  \int_{-\textrm{i}\infty}^{+\textrm{i}\infty} d\lambda 
  \exp\left[\lambda\left(\sum_{i=1}^N w_i -N \right) \right] \times 
  \nonumber \\
  &\times&
  \int_{0}^{+\infty} \prod_{\tau=1}^T d\mu_\tau
  \int_{-\textrm{i}\infty}^{+\textrm{i}\infty} \prod_{\tau=1}^T d\hat\mu_\tau
  \exp\left[ \sum_{\tau=1}^T \hat \mu_\tau 
    \left(u_\tau+v+\sum_{i=1}^N x_{i\tau} w_i-\mu_\tau \right) \right]
  \exp\left[ -\gamma (1-\beta)Tv-\gamma\sum_{\tau=1}^T u_\tau\right] \ \ \ ,
\end{eqnarray}
where the constraints are imposed by means of the Lagrange multipliers
$\lambda, \mu, \hat \mu$. In view of applying the trick in
(\ref{eq:replicatrick}), we introduce $n$ identical replicas of the system
corresponding to the same history of returns $\{x_{i\tau}\}$, and write down
$Z^n_\gamma[\{x_{i\tau}\}]$. After this, the average over the samples 
can be performed and allows one to introduce the \textsl{overlap} matrix
\begin{equation}
  Q^{ab} = \frac 1N \sum_{i=1}^N w_i^a w_i^b \ , 
  \quad a,b=1,\ldots n \ ,
  \label{eq:qab}
\end{equation}
as well as its conjugate $\hat Q^{ab}$ (the Lagrange multiplier imposing
(\ref{eq:qab})). Here, $a$ and $b$ are replica indexes. After (several)
Gaussian integrals, one is left with
\begin{eqnarray}
  \overline{Z^n_\gamma[\{x_{i\tau}\}]} 
  & \sim & 
  \int_{-\infty}^{+\infty} \prod_{a=1}^n dv^a    
  \int_{-\infty}^{+\infty} \prod_{a,b} dQ^{ab} 
  \int_{-\textrm{i}\infty}^{+\textrm{i}\infty} \prod_{a,b} d\hat Q^{ab}
  \exp \Bigg\{
  N\sum_{a,b} Q^{ab} \hat Q^{ab} - N \sum_{a,b} \hat Q^{ab} 
  - \gamma (1-\beta) T \sum_a v^a   \nonumber \\
  && 
  -Tn \log \gamma
  + T \log \hat Z_\gamma\left(\{v^a\},\{ Q^{ab}\}\right) 
  - \frac T2 \textrm{Tr} \log Q - \frac N2 \textrm{Tr} \log \hat Q 
  -\frac{nN}{2}\log 2 \Bigg\} \ , 
  \label{eq:zn}
\end{eqnarray}
where 
\begin{equation}
  \hat Z_\gamma\left(\{v^a\},\{Q^{ab}\}\right) 
  \equiv 
  \int_{-\infty}^{+\infty} \prod_{a=1}^n dy^a 
  \exp\left[
    - \frac 12 \sum_{a,b=1}^n 
    (Q^{-1})^{ab}(y^a-v^a) (y^b-v^b)
    +\gamma\sum_{a=1}^n y^a \theta(-y^a)
    \right] \ .
  \label{eq:zeff}
\end{equation}
We now write $T=tN$ and work at fixed $t$ while $N\to \infty$.

The most natural solution is obtained by realizing that all the replicas are
identical. Given the linear character of the problem, the symmetric solution
should be the correct one. The replica-symmetric solution corresponds to the
\emph{ansatz}
\begin{equation}
  Q^{ab} = 
  \begin{cases} 
    q_1 & \text{if $a=b$} \\
    q_0 & \text{if $a\neq b$}
  \end{cases}
  \ ;
  \quad
  \hat Q^{ab} = 
  \begin{cases} 
    \hat q_1 & \text{if $a=b$} \\
    \hat q_0 & \text{if $a\neq b$}
  \end{cases}
  \ ,
  \label{eq:ansatz}
\end{equation}
and $v^a=v$ for any $a$. As we discuss in detail in appendix \ref{sec:appA}, one can
show that the optimal cost function, computed as from eq.~(\ref{eq:epsxit}) but
with the average of the log, is the minimum of
\begin{equation}
  \eps(v,q_0,\Delta) = \frac 1{2\Delta} + \Delta 
  \left[
    t(1-\beta)v-\frac{q_0}{2} + \frac{t}{2\sqrt{\pi}}
    \int_{-\infty}^{+\infty} ds e^{-s^2} g(v+s\sqrt{2q_0})
    \right] \ , 
  \label{eq:egs}
\end{equation}
 where $\Delta \equiv \lim_{\gamma\to\infty} \gamma\Delta q$ and the
 function $g(\cdot)$ is defined as
\begin{equation}
  g(x) = 
  \begin{cases} 
    0 & x \ge 0 \ ,\\ x^2 & -1 \le x < 0 \ ,\\ -2x-1 & x < -1 \ .
  \end{cases}
\end{equation}
Note that this function and its derivative are continuous.  Moreover, $v$
and $q_0$ in (\ref{eq:egs}) are solutions of the saddle point equations
\begin{eqnarray}
  1-\beta+\frac{1}{2\sqrt{\pi}} \int ds e^{-s^2} g'(v+s\sqrt{2q_0}) 
  & = & 0 \ , \label{eq:spv}\\
  -1 + \frac{t}{\sqrt{2\pi q_0}} \int ds e^{-s^2} s\  g'(v+s\sqrt{2q_0})
  & = & 0 \label{eq:spq1}\ . 
\end{eqnarray}

We require that the minimum of (\ref{eq:egs}) occur at a finite value of
$\Delta$. In order to understand this point, we recall the meaning 
of $\Delta$ (see also (\ref{eq:ansatz})):
\begin{equation}
  \Delta/\gamma \sim \Delta q = (q_1-q_0) = \frac 1N \sum_{i=1}^N
  \big(w_i^{(1)}\big)^2 - \frac 1N \sum_{i=1}^N w_i^{(1)} w_i^{(2)} \sim
  \overline{w^2} -\overline{w}^2 \ ,
\end{equation}
where the superscripts (1) and (2) represent two generic replicas of the
system. We then find that $\Delta$ is proportional to the fluctuations in
the distribution of the $w$'s. An infinite value of $\Delta$ would then
correspond to a portfolio which is infinitely short on some particular
positions and, because of the global budget constraint~(\ref{eq:budget}),
infinitely long on some other ones.

Given (\ref{eq:egs}), the existence of a solution at finite $\Delta$
translates into the following condition:
\begin{equation}
  t(1-\beta)v-\frac{q_0}{2} + \frac{t}{2\sqrt{\pi}}
  \int_{-\infty}^{+\infty} ds e^{-s^2} g(v+s\sqrt{2q_0}) 
   \ge  0 \ , \label{eq:hastobepos}
\end{equation}
which defines, along with eqs.~(\ref{eq:spv}) and (\ref{eq:spq1}), our phase
diagram.

\section{The phase diagram \label{sec:phasediag}}


\begin{figure}
  \psfrag{N/T}[][][2.5]{$\displaystyle \frac NT$}
  \psfrag{b}[][][2.5]{$\beta$}
  \psfrag{Delta}[][][2.5]{$\Delta$}
  \psfrag{b = 0.95}[][][2.5]{$\beta=0.95$}
  \psfrag{b = 0.90}[][][2.5]{$\beta=0.90$}
  \psfrag{b = 0.80}[][][2.5]{$\beta=0.80$}
  \psfrag{1suradice}[][][2.5]{$1/\sqrt{x}$}
  \psfrag{ntmntc}[][][2.5]{$\displaystyle \left[\frac N T -\left(\frac
  NT\right)^*\right]$}
  \begin{center}
    \includegraphics[angle=-90,width=.49\textwidth]{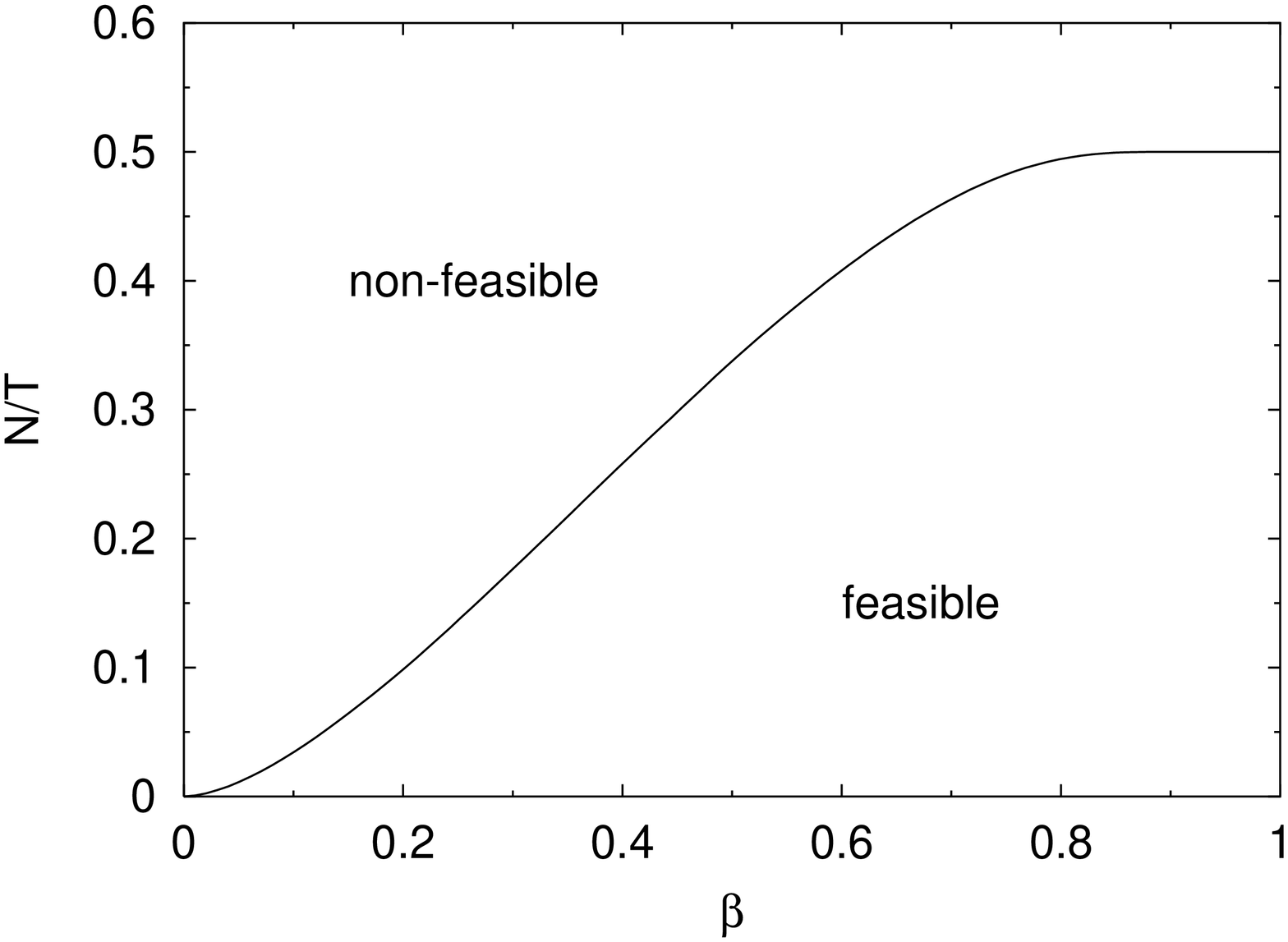}
    \includegraphics[angle=-90,width=.49\textwidth]{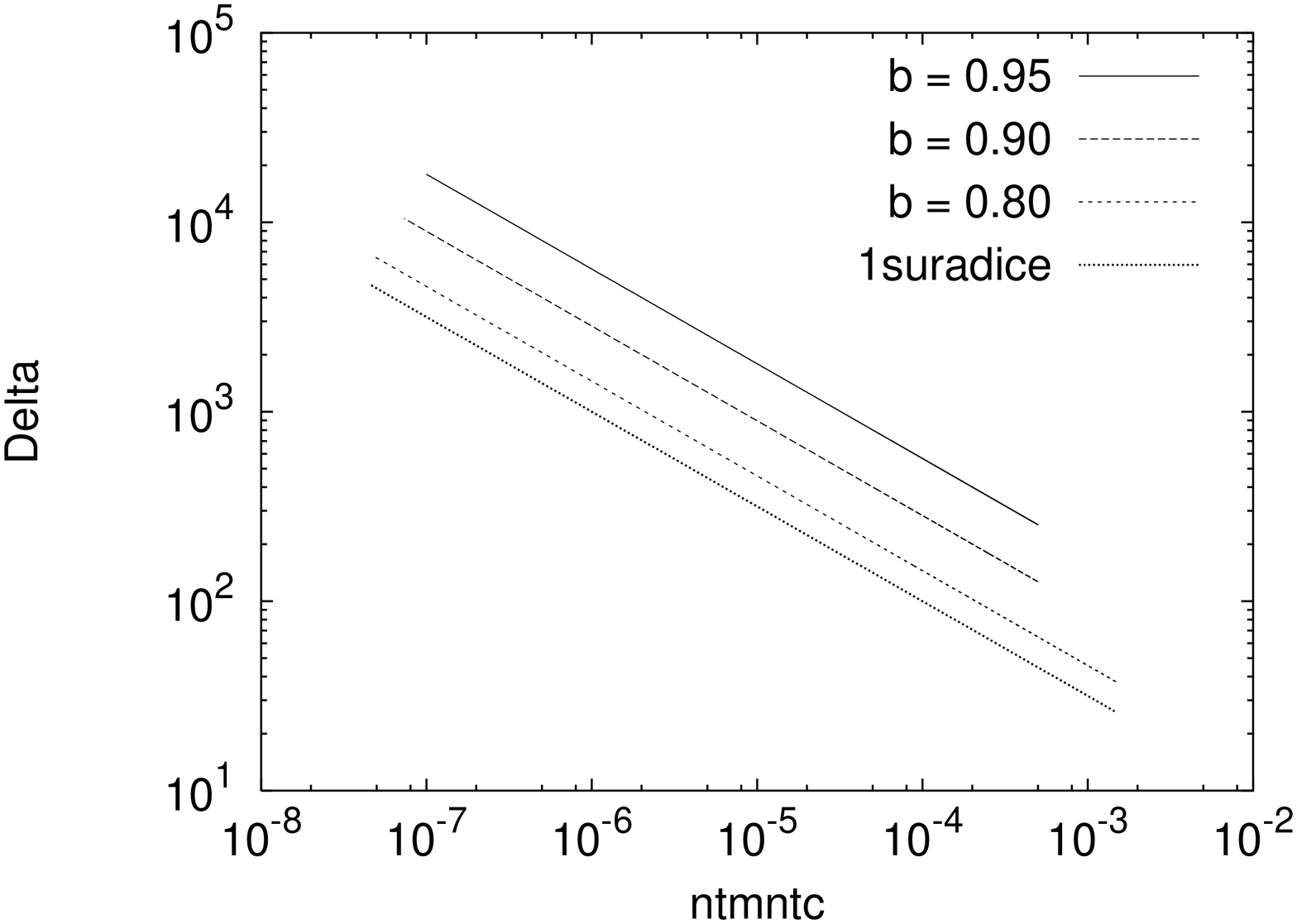}
  \end{center}
  \caption{Left panel: The phase diagram of the feasibility problem for
  expected shortfall. Right panel: The order parameter $\Delta$ diverges
  with an exponent $1/2$ as the transition line is approached. A curve of
  slope $-1/2$ is also shown for comparison. }
\label{fig:pd}
\end{figure}

We can now chart the feasibility map of the expected shortfall problem. We
will use as control parameters $N/T\equiv 1/t$ and $\beta$. The limiting
case $\beta\to 1$ can be worked out analytically and one can show that the
critical value $t^*$ is given by
\begin{equation}
  \frac 1 {t^*} = \frac 12 - 
  \mathcal{O}\left[(1-\beta)^3 e^{-\left(4\pi(1-\beta)^2\right)^{-1}}\right] \ .
  \label{eq:1tlimit}
\end{equation}
This limit corresponds to the over-pessimistic case of maximal loss, in
which the single worst loss contribute to the risk measure. The optimization
problem is the following:
\begin{equation}
  \min_{{\bf w}} \bigg[ \max_{\tau\in\{1,\ldots T\}}
    \Big(-\sum_i w_i x_{i\tau} \Big) \bigg] \ .
  \label{eq:maxloss}
\end{equation}
A simple ``geometric'' argument by Kondor et al.~\cite{kondor05} borrowed
from a random geometry context~\cite{rangeo} leads to the critical value
$1/t^*=0.5$ in this extreme case. The idea is the following. According to
eq.~(\ref{eq:maxloss}), one has to look for the minimum of a polytope made
by a large number of planes, whose normal vectors (the $x_{i\tau}$) are
drawn from a symmetric distribution. The simplex is convex, but with some
probability it can be unbounded from below and the optimization problem is
ill defined. Increasing $T$ means that the probability of this event
decreases, because there are more planes and thus it is more likely that for
large values of the $w_i$ the max over $t$ has a positive slope in the
$i$-th direction. The exact law for this probability can be obtained by
induction on $N$ and $T$~\cite{kondor05} and, as we said, it jumps in the
thermodynamic limit from 1 to 0 at $N/T=0.5$. Given that the corrections to
this limit case are exponentially small (eq.~(\ref{eq:1tlimit})), the
threshold $0.5$ can be considered as a good approximation of the actual
value for many cases of practical interest (i.e. $\beta \gtrsim 0.9$).

\begin{figure}
  \psfrag{ntmntc}[][][2.5]{$\displaystyle \left[\frac N T -\left(\frac
  NT\right)^*\right] \cdot N^{0.5}$}
  \psfrag{N/T}[][][2.5]{$\displaystyle \frac NT$}
  \psfrag{proba}[][][2.5]{probability}
  \psfrag{N=32}[][][2.5]{$N=32~~~~~~$}
  \psfrag{N=64}[][][2.5]{$N=64~~~~~~$}
  \psfrag{N=128}[][][2.5]{$N=128~~~$}
  \psfrag{N=256}[][][2.5]{$N=256~~~$}
  \psfrag{N=512}[][][2.5]{$N=512~~~$}
  \psfrag{N=1024}[][][2.5]{$N=1024$}
  \begin{center}
    \includegraphics[angle=-90,width=.49\textwidth]{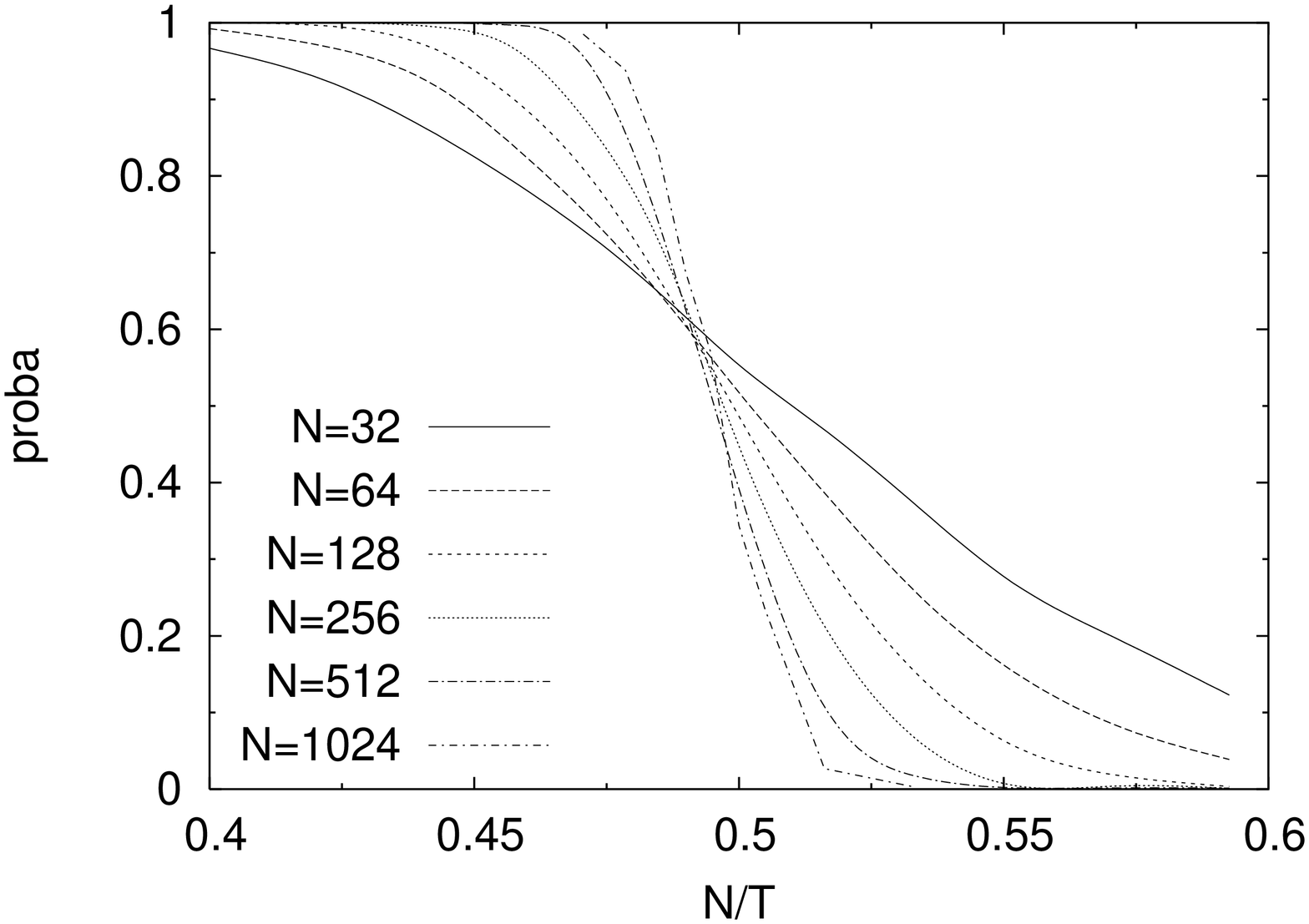}
    \includegraphics[angle=-90,width=.49\textwidth]{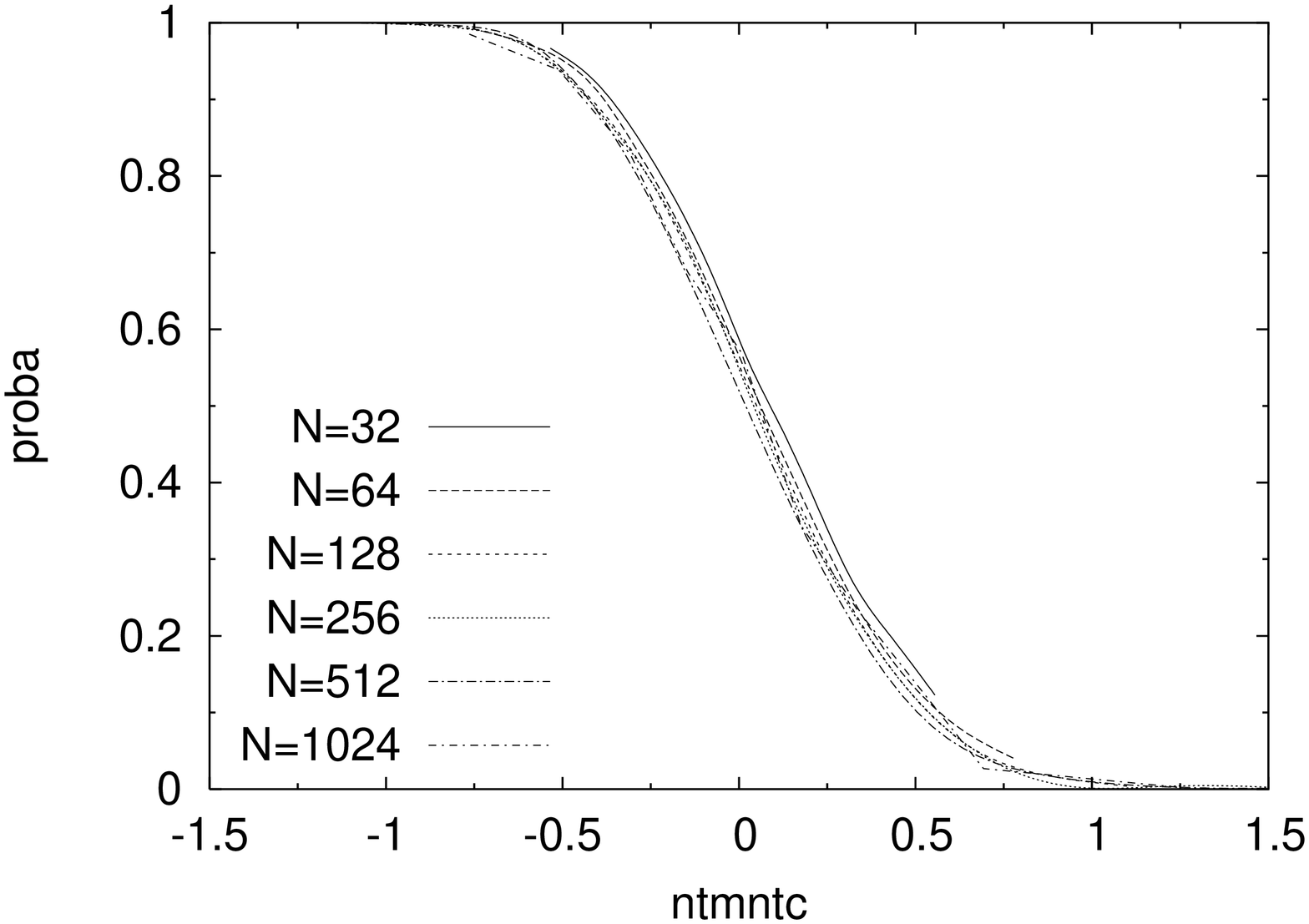}
  \end{center}
  \caption{Left: The probability of finding a finite solution as obtained
  from linear programming at increasing values of $N$ and with
  $\beta=0.8$. Right: Scaling plot of the same data. The critical value is
  set equal to the analytical one, $N/T = 0.4945$ and the critical exponent
  is $1/2$, i.e. the one obtained in \cite{kondor05} for the limit case
  $\beta\to 1$. The data do not collapse perfectly, and better results can
  be obtained by slightly changing either the critical value or the
  exponent.}
\label{fig:proba08}
\end{figure}

For finite values of $\beta$ we have solved numerically eqs.~(\ref{eq:spv}),
(\ref{eq:spq1}) and (\ref{eq:hastobepos}) using the following procedure. We
first solve the two equations (\ref{eq:spv}) and (\ref{eq:spq1}), which
always admit a solution for $(v,q_0)$. We then plot the l.h.s. of
eq.~(\ref{eq:hastobepos}) as a function of $1/t$ for a fixed value of
$\beta$. This function is positive at small $1/t$ and becomes negative
beyond a threshold $1/t^*$. By keeping track of $1/t^*$ (numerically
obtaining via linear interpolations) for each value of $\beta$ we build up
the phase diagram (Fig.~\ref{fig:pd}, left).  We show in the right panel of
Fig.~\ref{fig:pd} the divergence of the order parameter $\Delta$ versus $1/t
- 1/t^*$. The critical exponent is found to be $1/2$:
\begin{equation}
  \Delta \sim \left(\frac 1t - \frac 1{t^*(\beta)}\right)^{-1/2} \ .
\end{equation},
again in agreement with the scaling found in \cite{kondor05}.  We have
performed extensive numerical simulations in order to check the validity of
our analytical findings. For a given realization of the time series, we
solve the optimization problem (\ref{eq:ebeta}) by standard linear
programming~\cite{numrec}. We impose a large negative cutoff for the $w$'s,
that is $w_i>-W$, and we say that a feasible solution exists if it stays
finite for $W\to\infty$. We then repeat the procedure for a certain number
of samples, and then average our final results (optimal cost, optimal $v$,
and the variance of the $w$'s in the optimal portfolio) over those of them
which produced a finite solution. In Fig.~\ref{fig:proba08} we show how the
probability of finding a finite solution depends on the size of the problem.
Here, the probability is simply defined in terms of the frequency. We see
that the convergence towards the expected $1$-$0$ law is fairly slow, and a
finite size scaling analysis is shown in the right panel. Without loss of
generality, we can summarize the finite-$N$ numerical results by writing the
probability of finding a finite solution as
\begin{equation}
  p(N,T,\beta) = f\left[ \left(\frac 1t - \frac 1{t^*(\beta)}\right)\cdot
  N^{\alpha(\beta)}\right] \ ,
  \label{eq:pntb}
\end{equation}
where $f(x)\to 1$ if $x\gg 1$ and $f(x)\to 0$ if $x\ll 1$, and where
$\alpha(1)=1/2$. 

\begin{figure}
  \psfrag{N/T}[][][2.5]{$\displaystyle \frac NT$}
  \psfrag{energy}[][][2.5]{cost function}
  \psfrag{N=inf}[][][2.5]{analytic}
  \psfrag{optimal v}[][][2.5]{$v_\textrm{opt}$}
  \psfrag{N=32}[][][2.5]{$N=32~~~~$}
  \psfrag{N=64}[][][2.5]{$N=64~~~~$}
  \psfrag{N=128}[][][2.5]{$N=128$}
  \psfrag{N=256}[][][2.5]{$N=256$}
  \begin{center}
    \includegraphics[angle=-90,width=.49\textwidth]{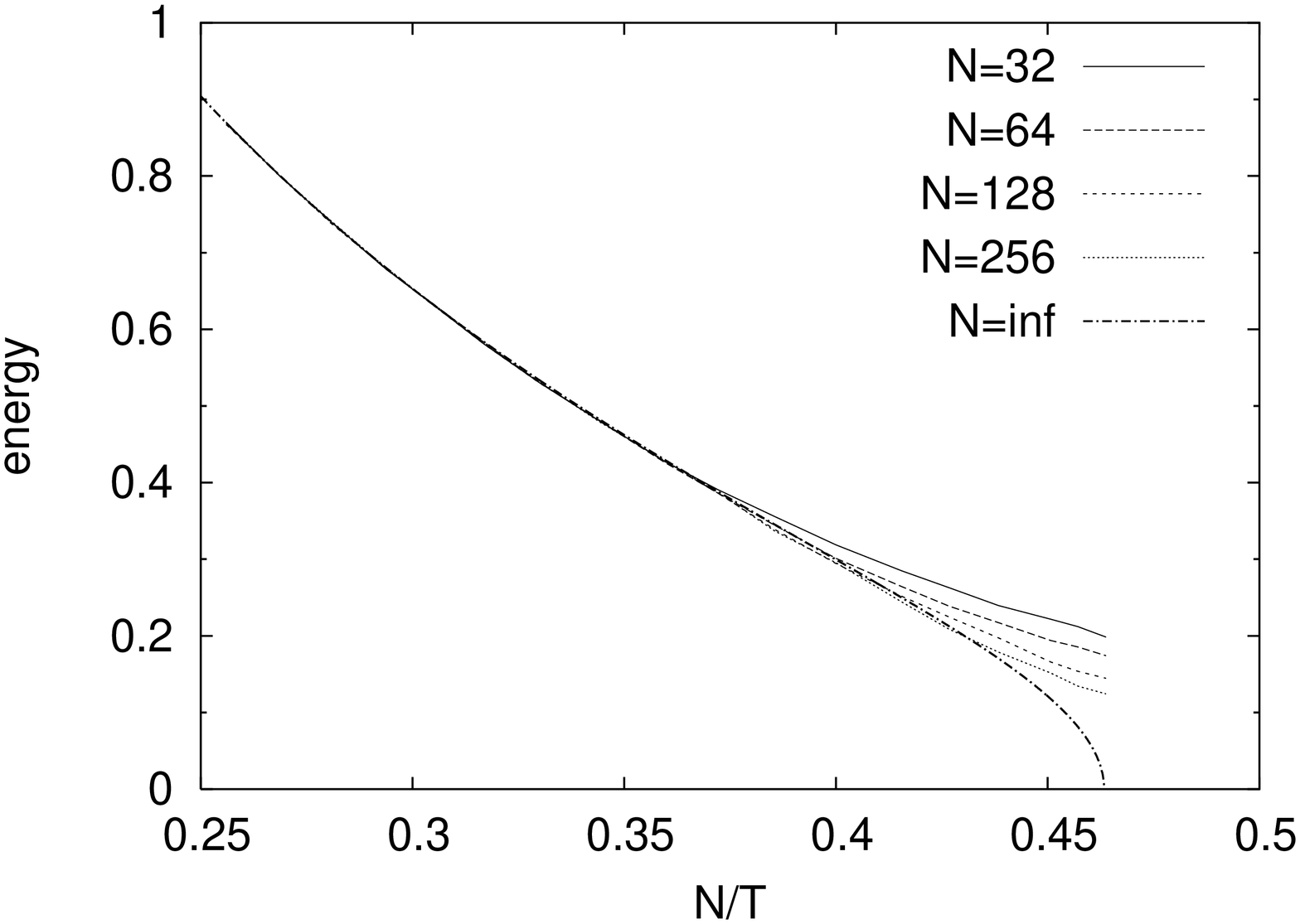}
    \includegraphics[angle=-90,width=.49\textwidth]{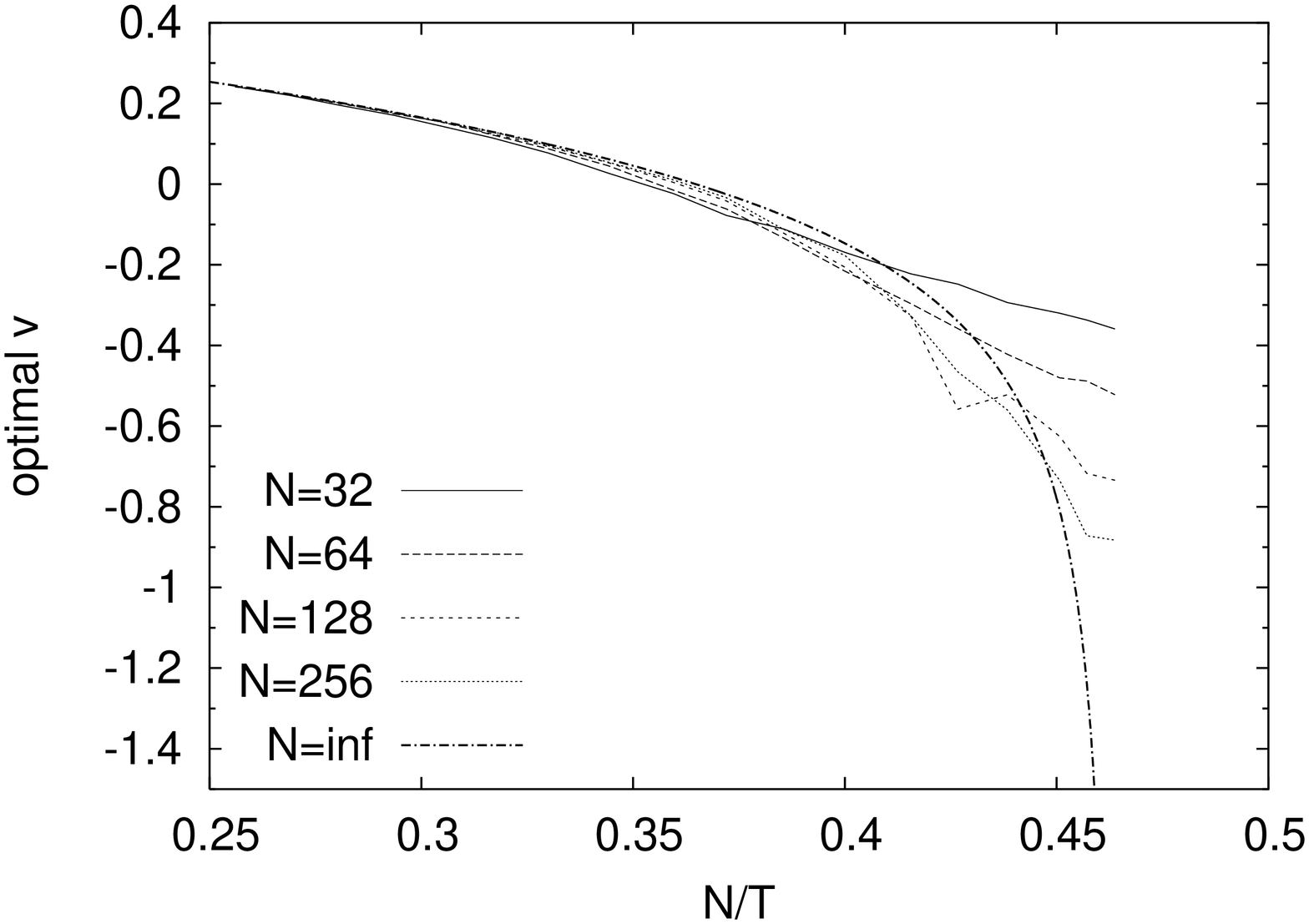}
  \end{center}
  \caption{Numerical results from linear programming and comparison with
  analytical predictions at large $N$. Left: The minimum cost of the
  optimization problem vs $N/T$, at increasing values of $N$. The thick line
  is the analytical solution (\ref{eq:egs}). Here $\beta=0.7$, $(N/T)^*
  \simeq 0.463$. Right: The optimal value of $v$ as found numerically for
  several values of $N$ is compared to the analytical solution. }
\label{fig:energy}
\end{figure}

In Fig.~\ref{fig:energy} (left panel) we plot, for a given value of $\beta$,
the optimal cost found numerically for several values of the size $N$
compared to the analytical prediction at infinite $N$. One can show that the
cost vanishes as $\Delta^{-1}\sim (1/t-1/t^*)^{1/2}$. The right panel of
the same figure shows the behavior of the value of $v$ which leads to the
optimal cost versus $N/T$, for the same fixed value of $\beta$. Also in
this case, the analytical ($N\to \infty$ limit) is plotted for comparison.
We note that this quantity was suggested~\cite{rockafellar00} to be a good
approximation of the VaR of the optimal portfolio: We find here that
$v_\textrm{opt}$ diverges at the critical threshold and becomes negative at
an even smaller value of $N/T$.

\section{Conclusions \label{sec:conclu}}

We have shown that the problem of optimizing a portfolio under the expected
shortfall measure of risk by using empirical distributions of returns is not
well defined when the ratio $N/T$ of assets to data points is larger than a
certain critical value. This value depends on the threshold $\beta$ of the
risk measure in a continuous way and this defines a phase diagram. The lower
the value of $\beta$, the larger the length of the time series needed for
the portfolio optimization. The analytical approach we have discussed in
this paper allows us to have a clear understanding of this phase
transition. The mathematical reason for the non-feasibility of the
optimization problem is that, with a certain probability $p(N,T,\beta)$, the
linear constraints in (\ref{eq:constraints}) define a simplex which is not
bounded from below, thus leading to a solution which is not finite ($\Delta
q \to\infty$ in our language), in the same way as it happens in the extreme
case $\beta\to 1$ discussed in \cite{kondor05}. From a more physical point
of view, it is reasonable that the feasibility of the problem depend on the
number of data points we take from the time series with respect to the
number of financial instruments of our portfolio. The probabilistic
character of the time series is reflected in the probability
$p(N,T,\beta)$. Interestingly, this probability becomes a threshold function
at large $N$ if $N/T\equiv 1/t$ is finite, and its general form is given in
(\ref{eq:pntb}).

These results have a practical relevance in portfolio optimization. The
order parameter discussed in this paper is tightly related to the relative
estimation error~\cite{pafka02}. The fact that this order parameter has been
found to diverge means that in some regions of the parameter space the
estimation error blows up, which makes the task of portfolio optimization
completely meaningless.  The divergence of estimation error is not limited
to the case of expected shortfall. As shown in \cite{kondor05}, it happens
in the case of variance and absolute deviation as well \footnote{see also
S.~Ciliberti et al., Proceeding of the Torino Conference on ``Applications
of Physics in Financial Analysis'' (2006), for a replica approach to the
portfolio optimization problem under absolute deviation}, but the noise
sensitivity of expected shortfall turns out to be even greater than that of
these more conventional risk measures.

There is nothing surprising about the fact that if there are no sufficient
data, the estimation error is large and we cannot make a good decision. What
is surprising is that there is a sharply defined threshold where the
estimation error actually diverges.

For a given portfolio size, it is important to know that a minimum amount of
data points is required in order to perform an optimization based on
empirical distributions.  We also note that the divergence of the parameter
$\Delta$ at the phase transition, which is directly related to the
fluctuations of the optimal portfolio, may play a dramatic role in practical
cases. To stress this point, we can define a sort of ``susceptibility'' with
respect to the data,
\begin{equation}
  \chi^\tau_{ij} = \frac{\partial \langle w_j \rangle}{ \partial x_{i\tau}} \ ,
\end{equation}
and one can show that this quantity diverges at the critical point, since
$\chi^\tau_{ij} \sim \Delta$. A small change (or uncertainty) in $x_{i\tau}$
becomes increasingly relevant as the transition is approached, and the
portfolio optimization could then be very unstable even in the feasible
region of the phase diagram. We stress that the susceptibility we have
introduced might be considered as a measure of the effect of the noise on
portfolio selection and is very reminiscent to the measure proposed
in~\cite{pafka02}.

In order to present a clean, analytic picture, we have made several
simplifying assumptions in this work. We have omitted the constraint on the
returns, liquidity constraints, correlations between the assets,
nonstationary effects, etc. Some of these can be systematically taken into
account and we plan to return to these finer details in a subsequent work.

\emph{Acknowledgments.} We thank O.\ C.\ Martin, and M.\ Potters
for useful discussions, and particularly J.\ P.\ Bouchaud for a critical
reading of the manuscript. S.~C. is supported by EC through the network MTR
2002-00319, STIPCO, I.K. by the National Office of Research and 
Technology under grant No. KCKHA005.

\appendix

\section{The replica symmetric solution \label{sec:appA}}

We show in this appendix how the minimum cost function corresponding to the
replica-symmetric ansatz is obtained.

The `TrLog$Q$' term in (\ref{eq:zn}) is computed by realizing that the
eigenvalues of such a symmetric matrix are $(q_1+(n-1)q_0)$ (with multiplicity
$1$) and $(q_1-q_0)$ with multiplicity $n-1$. Then,
\begin{equation}
  \textrm{Tr} \log Q = \log \textrm{det} Q = 
  \log(q_1+(n-1)q_0) + (n-1)\log(q_1-q_0) =
  n\left(\log\Delta q + \frac {q_1}{\Delta q}\right)
  + \mathcal{O}(n^2)\ ,
\end{equation}
where $\Delta q \equiv q_1-q_0$.  
The effective partition function in (\ref{eq:zeff}) depends on $Q^{-1}$,
whose elements are:
\begin{equation}
   (Q^{-1})^{ab} = 
  \begin{cases} 
     (\Delta q -q_0)/(\Delta q)^2  + \mathcal{O}(n)
    & \text{if $a= b$} \\
    - q_0/(\Delta q)^2  + \mathcal{O}(n)
    & \text{if $a\neq b$}      
  \end{cases}
\end{equation}
By introducing a Gaussian measure $dP_{q_0}(s)\equiv \frac{ds}{\sqrt{2\pi q_0}}
e^{-s^2/2 q_0}$, one can show that
\begin{eqnarray}
  \frac 1n \log\hat Z(v, q_1, q_0) & = & 
  \frac 1n 
  \log \left\{
  \int \prod_a dx_a 
  e^{-\frac 1{2\Delta q} \sum_a (x^a)^2 + 
    \gamma \sum_a (x^a+v)\theta(-x^a-v)} 
  \int dP_{q_0}(s) e^{\frac{s}{\Delta q} \sum_a x^a} 
  \right\}
  \nonumber \\
  & = & 
  \frac{q_0}{2\Delta q} + 
  \int dP_{q_0}(s) \log B_\gamma(s,v,\Delta q) 
  + \mathcal{O}(n)
\end{eqnarray}
where we have defined
\begin{eqnarray}
  B_\gamma(s,v,\Delta q) & \equiv & 
  \int dx \exp\left(
  -\frac{(x-s)^2}{2\Delta q} + \gamma(x+v)\theta(-x-v)
  \right) 
  \ .
\end{eqnarray}

The exponential in (\ref{eq:zn}) now reads exp$Nn[S(q_0,\Delta q, \hat q_0,
\Delta\hat q) + \mathcal{O}(n) ]$, where
\begin{equation}
  \begin{split}
  S(q_0,\Delta q, \hat q_0,\Delta\hat q)=
  q_0 \Delta \hat q + \hat q_0 \Delta q +\Delta q\Delta\hat q - 
  \Delta\hat q -\gamma t(1-\beta)v -t \log\gamma + 
  t\int dP_{q_0}(s) \log B_\gamma(s,v,\Delta q) \\
  - \frac t2\log\Delta q 
  - \frac 12\left(\log\Delta \hat q + \frac{\hat q_0}{\Delta \hat q}\right)-
  \frac{\log 2}{2} \ .    
  \end{split}
  \label{eq:exp0}
\end{equation}
The saddle point equations for $\hat q_0$ and $\Delta \hat q$ allow then to
simplify this expression. The free energy $(-\gamma) f_\gamma = \lim_{n\to
0}\partial \overline{Z^n_\gamma} /\partial n$ is given by
\begin{equation}
  -\gamma f_\gamma(v,q_0,\Delta q) = \frac 12-t\log \gamma
  +\frac{1-t}{2}\log\Delta q + \frac{q_0-1}{2\Delta q} - \gamma t(1-\beta) v
  + t\int dP_{q_0}(s) \log B_\gamma(s,v,\Delta q) \ ,
  \label{eq:freeen}
\end{equation}
where the actual values of $v,q_0$ and $\Delta q$ are fixed by the saddle
point equations 
\begin{equation}
  \frac{\partial f_\gamma}{\partial v} = 
  \frac{\partial f_\gamma}{\partial q_0} = 
  \frac{\partial f_\gamma}{\partial \Delta q} = 0 \ . 
\end{equation}
A close inspection of these saddle point equations allows one to perform the low
temperature $\gamma\to\infty$ limit by assuming that $\Delta
q=\Delta/\gamma$ while $v$ and $q_0$ do not depend on the temperature. In
this limit one can show that
\begin{equation}
  \lim_{\gamma\to\infty}
  \frac 1\gamma \log B_\gamma(s,v,\Delta/\gamma) = 
  \begin{cases}
    s+v+\Delta/2 & \quad s<-v-\Delta\\
    -(v+s)^2/2\Delta & \quad -v-\Delta \le s < -v \\
    0 & \quad s\ge -v
  \end{cases}
\end{equation}
If we plug this expression into eq.~(\ref{eq:freeen}) and perform the
large-$\gamma$ limit we get the minimum cost:
\begin{equation}
  E = \lim_{\gamma\to\infty} f_\gamma = -\frac{q_0-1}{2 \Delta } + 
  t(1-\beta) v - t\int_{-\infty}^{-\Delta} \frac {dx}{\sqrt{2\pi q_0}}
  e^{-\frac{(x-v)^2}{2q_0}} \left(x+\frac\Delta 2\right) +
  \frac{t}{2\Delta} \int_{-\Delta}^{0} \frac {dx}{\sqrt{2\pi q_0}}
  e^{-\frac{(x-v)^2}{2q_0}} x^2 \ .
\end{equation}
We rescale $x\to x\Delta$, $v\to v\Delta$, and $q_0\to q_0\Delta^2$, and
after some algebra we obtain eq.~(\ref{eq:egs}).

\end{document}